\documentclass[12pt]{iopart}

\usepackage{graphicx}
\usepackage{iopams}

\begin{document}

\title[Spiral wave dynamics in a neuronal network model]
{Spiral wave dynamics in a neuronal network model}

\author{Diogo L. M. Souza$^{1,*}$, Fernando S. Borges$^{2,3}$, Enrique C. Gabrick$^{1,4,5}$, Lucas E. Bentivoglio$^1$, Paulo R. Protachevicz$^6$, Vagner dos Santos$^{8,9}$, Ricardo L. Viana$^7$, Ibere L. Caldas$^6$, Kelly C. Iarosz$^{1,6,8}$, Antonio M. Batista$^{1,6,10}$, Jürgen Kurths$^{4,5}$}

\address{$^1$ \quad Graduate Program in Sciences, State University of Ponta Grossa, 84030-900, Ponta Grossa, PR, Brazil \\
$^2$ \quad Department of Physiology and Pharmacology, State University of New York Downstate Health Sciences University, Brooklyn, New York, USA \\
$^3$ \quad Center for Mathematics, Computation, and Cognition, Federal University of ABC, 09606-045 S\~ao Bernardo do Campo, SP, Brazil \\
$^4$ \quad Department of Physics, Humboldt University Berlin, Newtonstra\ss e 15, 12489 Berlin, Germany \\
$^5$ \quad Potsdam Institute for Climate Impact Research, Telegrafenberg A31, 14473 Potsdam, Germany \\
$^6$ \quad Institute of Physics, University of S\~ao Paulo, 05508-090, S\~ao Paulo, SP, Brazil \\
$^7$ \quad Department of Physics, Federal University of Paran\'a, 81530-000, Curitiba, PR, Brazil \\
$^8$ \quad Exact and Natural Sciences and Engineering, UNIFATEB University Center, 84266-010, Telêmaco Borba, PR, Brazil.\\
$^9$ \quad Department of Physics, State University of Ponta Grossa, 84030-900, Ponta Grossa, PR, Brazil\\
$^{10}$ \quad Department of Mathematics and Statistics, State University of Ponta Grossa, 84030-900, Ponta Grossa, PR, Brazil}

\ead{diogoleonaisouza@gmail.com}

\begin{abstract}
Spiral waves are spatial-temporal patterns that can emerge in different systems as heart tissues, chemical oscillators, ecological networks and the brain. These waves have been identified in the neocortex of turtles, rats, and humans, particularly during sleep-like states. Although their functions in cognitive activities remain until now poorly understood, these patterns are related to cortical activity modulation and contribute to cortical processing. In this work, we construct a neuronal network layer based on the spatial distribution of pyramidal neurons. Our main goal is to investigate how local connectivity and coupling strength are associated with the emergence of spiral waves. Therefore, we propose a trustworthy method capable of detecting different wave patterns, based on local and global phase order parameters. As a result, we find that the range of connection radius ($R$) plays a crucial role in the appearance of spiral waves. For $R<20$ $\mu$m, only asynchronous activity is observed due to small number of connections. The coupling strength ($g_{\rm syn}$) greatly influences the pattern transitions for higher $R$, where spikes and bursts firing patterns can be observed in spiral and non-spiral waves. Finally, we show that for some values of $R$ and $g_{\rm syn}$ bistable states of wave patterns are obtained.

\end{abstract}

\vspace{2pc}
\noindent{\it Keywords}: firing patterns, spiral waves, network, neural oscillations, chimera, pyramidal neurons

%%%%%%%%%%%%%%%%%%%%%%%%%%%%%%%%%%%%%%%%%%
%%%%%%%%%%%%%%%%%%%%%%%%%%%%%%%%%%%%%%%%%%

\section{Introduction}

It has been identified astonishing spatiotemporal patterns which can be seen in 2D and 3D networks with non-local interactions, named spiral wave chimera states. It consists of synchronous oscillators which rotate around a desynchronized core, which is called phase singularity (PS) \cite{chaoli2018}. This pattern is seen in a variety of fields. In heart tissues \cite{ding2023, holden1997, suran2019} spiral waves have been extensively studied due their correlation with some heart issues, such as ventricular tachycardia and ventricular fibrillation. Davidenko et al. \cite{Davidenko1992} demonstrated the emergence of spiral waves in sheep and dog epicardial muscle by means of potentiometric dye with charge-coupled device imaging technology. The authors showed that sometimes the phase singularity drifted away from its original position and dissipated in the tissue border, this drift was associated with a Doppler shift. Supression of spiral waves in the atrial cardiomyocytes can be done by means of optogenetics \cite{Bingen2014}. Spiral waves are able to emerge in ecological network composed of diffusible prey-predator species locally coupled \cite{kundu2018}. Totz et al. \cite{totz2018} observed the emergence of spiral wave chimeras in large populations of coupled chemical oscillators, they have studied the motion and splitting of the asynchronous core. 

The chimera is defined as the coexistence of coherent and incoherent domains when similar elements interact \cite{Glaze2016}. The notion of chimera states has been discussed in the last decade \cite{Haugland2021}. Majhi et al. \cite{Majhi2019} reviewed important aspects of such dynamics and highlighted the different types, as well as their relevance in a biophysical context. Chimera states play an important role in empirical brain networks, specially in the cerebral cortex \cite{Wang2020}.  It has been found in networks composed of coupled neurons, such as integrate-and-fire \cite{Santos2021,Provata2020}, Hodgkin-Huxley \cite{Glaze2016}, Hindmarsh-Rose \cite{Santos2017}, FitzHugh-Nagumo \cite{Zhang2022,Ramados2021},  Morris-Lecar \cite{Wu2013}, Nekorkin maps \cite{Burk2018} and others. The spatiotemporal patterns, in which coherence coexists with incoherence, have been reported in local \cite{Santos2019,VagnerSantos2020} and non-local network configurations \cite{Hizanidis2014,Tsigkri-DeSmedt2015}.

Many studies have improved the understanding of chimera states in neuronal networks. The parameter regions for chimera states in a two coupling configuration of identical thermally sensitive Hodgkin-Huxley neurons was characterized in Ref. \cite{Glaze2016}. Majhi \cite{Majhi2016} investigated chimera states in uncoupled neurons induced by multilayer interactions. They found that the competition between electrical and chemical synapses can lead to chimeras. H\"ovel et al. \cite{Hovel2016} studied the coexistence of chimera and traveling waves, as well as multi-stability, in the parameter space of a neuronal system of excitability type I. Chemical and electrical synapses can affect the features of chimera states and traveling waves \cite{Calim2020}. The coexistence of patterns can emerge due to electrical and chemical coupling in a multi-weighted of the memristive Fitzhugh-Nagumo network \cite{Ramados2021}. It was demonstrated the importance of such interactions to obtain synchronous and desynchronous behaviours coexisting \cite{Hussain2021}. Chimera states were also reported in time-delay networks \cite{Lucchetti2021,Nikitin2022}.

Neuronal networks can exhibit spiral wave spatial patterns when in a sleep-like state \cite{huang2004, huang2010}. It is able to modify ongoing activity in the cortex \cite{huang2010, nature2023}. Local excitatory interactions are of great importance in the generation of spiral waves \cite{huang2010}. The role of spiral waves in the neocortex is not yet well understood, thus simulations and experimental methods are crucial to better understand spiral wave dynamics. The emergence of spiral waves in computational simulations has been observed in different models throughout the time. Ma et al. \cite{ma2012} studied the robustness of the spiral wave when noise is applied to ion channels of the standard Hodking-Huxley model. They showed that the spiral wave does not sustain its activity, whether the noise increases or the fraction of active channels is below a threshold. The transmission delay enhances the coherence of spiral waves in a noisy network \cite{wang2008}. Spiral waves can be seen in non-locally coupled maps for some coupling parameters \cite{bukh2019}. Santos et. al. \cite{santos2021} demonstrated that chimera spiral waves with multi-cores appear in regular and fractal networks. Their study showed that the initial condition of the adaptation current has a crucial role in the emergence of spiral waves. 

The formation of spiral waves are not only observed in neuronal simulations, but also in experimental data \cite{uzelac2022, townsend2015, prechtl1997}. The effects of spiral waves in cognitive activity are not well defined, although it influences the frequencies \cite{huang2010}. Prechtl et al. \cite{prechtl1997} used visual stimuli to register the appearance of spiral waves in the turtle cortex. Furthermore, Huang et al. \cite{huang2004} captured the emergence of spiral waves in rat cortex and its influence in coordinating oscillations of neuronal populations. Spiral waves have been identified in the neocortex during sleep-like states and pharmacologically induced oscillations \cite{huang2010}. It is theorized that such spatial pattern plays an important role in organizing and modulating healthy and pathological brain activities, for instance, epilepsy \cite{huang2004, huang2010, prechtl1997}.  

In this work, we propose a neuronal network composed of pyramidal cells, that are identified as excitatory neurons. Pyramidal neurons are the most common cells in the brain, and the name is due to its characteristic shape \cite{Bekkers2011}. This neuron type is also the main one of the excitatory neuron family in the brain \cite{Bekkers2011}. Our network model is inspired by a two dimensional slice of pyramidal layer of the hippocampus \cite{Bezaire2016}. The model considered to describe a single neuron membrane potential is the adaptive exponential integrate-and-fire (aEIF) \cite{brette05}. Besides the simplicity of this model and its low computational cost, it exhibits a great biophysical accuracy \cite{neuronalDynamics2014, Hertag2012}. Due to the lack of reliable tools able to characterize chimera spiral waves, we propose a method to identify chimeras spiral wave, and analyze their dependence on the coupling radius ($R$) and the synapse coupling strength ($g_{\rm syn}$). Our results show that the spiral waves emerge for different combinations of $R$ and $g_{\rm syn}$. We observe spiral waves when the value of the global phase order parameter is low ($Z_{\rm g}<0.7$), the local phase order parameter value is high ($Z_{\rm L} \approx 1$) and the amount of phase singularity is not greater than a maximum value ($P_{\rm max} \leq 20$). We observe for low $R$ values the desynchronized patterns.
Moreover, there are many pattern transitions and bistable regions in which spiral and non-spiral waves are possible for higher $R$ values.

The paper is organized as follows: In Section 2, we present our neuronal network model. In Section 3, we show the diagnostic tools. Section 4 presents our results. In the last Section, we draw our conclusions.

%%%%%%%%%%%%%%%%%%%%%%%%%%%%%%%%%%%%%%%%%%
%%%%%%%%%%%%%%%%%%%%%%%%%%%%%%%%%%%%%%%%%%
\section{Neuronal Network Model}

We consider a network composed of coupled adaptive exponential integrate-and-fire (aEIF) neurons \cite{brette05}. The dynamics of the neuronal network is given by
\begin{eqnarray}\label{eqIF}
C \frac{d V_i}{d t} & = & - g_{\rm L} (V_i - E_{\rm L}) + g_{\rm L} {\Delta}_{\rm T} \exp \left(\frac{V_i - V_{\rm T}}{{\Delta}_{\rm T}} \right) -w_i +I 
\\ & + &(V_{\rm rev}-V_i)\sum_{j=1}^{N} g_j M_{ij} , \nonumber \\
\tau_w \frac{d w_i}{d t} & = & a (V_i - E_{\rm L}) - w_i,  \\
\tau_g \frac{d g_i}{dt} & = & -g_i. 
\end{eqnarray}
where $V_i$ is the membrane potential, $w_i$ is the adaptation variable, $g_i$ is the synaptic conductance of the neuron $i$, $I$ is the injected current, $C$ is the membrane capacitance, $g_{\rm L}$ is the leak conductance, $E_{\rm L}$ is the resting potential, $\Delta_{\rm T}$ is the slope factor, $V_{\rm T}$ is the threshold potential, $V_{\rm rev}$ is the potential reversal, $\tau_w$ is the adaptation time constant, $a$ is the level of subthreshold adaptation and $\tau_g$ is the synaptic time constant. The parameter $\Delta_{\rm T}$ controls the sharpness of the initial phase of the spike \cite{badel08}. This model improves the standard leaky integrate-and-fire by adding the adaptation current and the exponential term in the membrane potential. The adaptation current describes the slow activation and deactivation of some potassium ionic channels \cite{Brette2015}. Besides that, a spike threshold mechanism represented by the exponential term describes the fast arising when a action potential is generated \cite{neuronalDynamics2014, brette05, firingAdex}. For $\Delta_{\rm T}\rightarrow 0$, the neuron model becomes a standard leaky integrate-and-fire neuron model \cite{clopath07}. If $V_i$ reaches the threshold $V_{\rm{peak}}$ at a certain time $t$, the following reset conditions are applied: $V_i\rightarrow V_{\rm r}$,
$w_i\rightarrow w_{\rm r}=w+b$ and $g_i \rightarrow g_i+g_{\rm r}$. In our simulations, we consider the parameter values in Table \ref{tab1}. We constructed a two-dimensional excitatory neuronal network with 1000 $\mu$m X 1000 $\mu$m dimensions and $N = 17,324$ neurons in which the distance between each neuron is 7 $\mu$m in $x$-axis and 8 $\mu$m in $y$-axis \cite{Bezaire2016}. The neuron $i$ is connected with its neighbors that are within a radius $R$ ($M_{ij}=1$), as displayed in Figure \ref{fig1}(a).

\begin{table}[htb]
\begin{center}	
\caption{Description and values of the parameters in the AEIF system used in the simulations \cite{naud08}.}
\label{tab1}
\begin{tabular}{c l l}
\hline
\small \textbf{Parameter} & \small \textbf{Description} &\small \textbf{Value}\\ 	
\hline
\small $C$ &\footnotesize Membrane capacitance &\small 200 pF \\
\small $I$ &\footnotesize Constant input current & \small 500 pA \\ 
\small $N$ &\footnotesize Number of aEIF neurons & \small 17.324 neurons \\
\small $a$ &\footnotesize Subthreshold adaptation &\small 2.0 nS \\ 
\small $b$ &\footnotesize Triggered adaptation &\small 70 pA \\ 
\small $\rho$ &\footnotesize Superficial density of neurons & \small 0.018 neurons/cm$^2$ \\
\small $R$ &\footnotesize Radius connection & \small [10,80] $\mu$m \\ 
\small $M_{ij}$ &\footnotesize Adjacency matrix elements &\small 0 or 1 \\
\small $w_i(0)$ &\footnotesize Initial membrane potential &\small [0, 70] pA \\
\small $V_i(0)$ &\footnotesize Initial membrane potential &\small [-70, -45] mV \\
\small $V_{\rm T}$	&\footnotesize Potential threshold & \small-50 mV \\ 
\small $V_{\rm rev}$ & \footnotesize Reversal potential &\small 0 mV \\
\small $V_{\rm peak}$ &\footnotesize maximum potential &\small -40 mV \\ 
\small $V_{\rm r}$ &\footnotesize Reset potential & \small -58 mV \\ 
\small $E_{\rm L}$ &\footnotesize Leak reversal potential &\small -70 mV \\
\small $g_{\rm L}$ &\footnotesize Leak conductance & \small 12 nS \\ 
\small $g_{\rm syn}$ &\footnotesize Excitatory synaptic conductance & \small [0.0,0.25] nS \\
\small $t_{\rm ini}$ &\footnotesize Initial time in the analyses &\small 25 s \\ 
\small $t_{\rm fin}$ &\footnotesize Final time in the analyses &\small 30 s \\ 
\small $\tau_w$ & \footnotesize Adaptation time constant & \small 300 ms\\
\small $\tau_{\rm s}$ & \footnotesize Synaptic time constant &\small 2.728 ms \\ 
\small $\Delta_{\rm T}$	&\footnotesize Slope factor &\small 2 mV \\
\small $d_{\rm x}$ &\footnotesize {\it x}-axis distance  & \small 7 $ \mu$m \\
\small $d_{\rm y}$ &\footnotesize {\it y}-axis distance  & \small 8 $ \mu$m \\  	
\hline
\end{tabular}
\end{center}
\end{table}

\section{Diagnostic tools}

To identify spiking and bursting firing patterns, we utilize the mean coefficient of variation (CV) of the neuronal inter-spike interval (ISI) \cite{Borges2017, Protachevicz2019}, that is given by
\begin{eqnarray}
{\rm CV}=\frac{{\sigma}_{\rm{ISI}}}{\rm{\overline{ISI}}},
\label{CV}
\end{eqnarray}
where ${\sigma}_{\rm{ISI}}$ is the standard deviation of the ISI normalised by the mean $\overline{\rm ISI}$ \cite{gabbiani98}. The ISI is defined by ${\rm ISI}_m = t_{\rm m+1} - t_{\rm m}$, where $\rm t_{m}$ is the time of the $m$-th neuronal spike \cite{Protachevicz2018}. Spike pattern produces ${\rm CV} < 0.5$, meanwhile, burst pattern produces ${\rm CV} \geq 0.5$ \cite{Protachevicz2019}. The mean firing rate is calculated by means of $F = \overline{\rm ISI}^{-1}$ \cite{Protachevicz2019}.

To determine synchronous behaviour of the whole network, we consider the global phase order parameter, that is defined as \cite{Kuramoto84}
\begin{equation}
Z(t)= \left|\frac{1}{N}\sum_{j=1}^{N}\exp({\rm i}\phi_{j}(t))\right|,
\end{equation}
where $\phi_j$ is the $j$-neuron phase, given by
\begin{equation}
\phi_{j}(t)=2\pi m+2\pi\frac{t-t_{j}^{m}}{t_{j}^{m+1}-t_{j}^{m}},
\end{equation}
$t_{j}^{m}$ corresponds to the time when a $m$-th spike ($m=0,1,2,\dots$) of a neuron $j$ happens ($t_{j}^{m} < t < t_{j}^{m+1}$). When the network is totally synchronized $Z(t) = 1$, total desynchronized states produce $Z(t) = 0$ and when $0 < Z(t) < 1$ the network is partially synchronized \cite{Kuramoto84, Protachevicz2018}. To better understanding of the network dynamics, we calculate the time-average of the order-parameter, that is given by
\begin{equation}
Z_{\rm g}=\frac{1}{t_{\rm fin}-{t_{\rm ini}}}\int_{t_{\rm ini}}^{t_{\rm fin}}Z(t) dt,
\end{equation} 
where $t_{\rm fin}-t_{\rm ini}$ is the time window. We use a time window equal to 5 s to compute ${\overline Z}$ to avoid transient effects.

We consider the local phase order parameter to study the local synchronization of each neuron. We separate the 1000 X 1000 $\mu$m network in 25 boxes in each dimension, totalizing 625 boxes, then each box contains a significant amount of neurons. The boxes are labeled by the subindex $(i, j)$, where $i$ indicates the $x$-axis and $j$ the $y$-axis position of the box, $i,j = [1,25]$. The size of each box is $40 \times 40$ $\mu$m and each one of them contains approximately 28 neurons. The local phase order parameter is given by \cite{Santos2020}
\begin{equation}
Z_{i,j} (t)=\left | \left[ \frac{A^2}{N^2} \right] \sum_{m=X_i}^{X_i+40} \sum_{n=Y_j}^{Y_{j}+40}\exp({\rm k}\phi_{m, n})\right|,
\end{equation}
where $k=\sqrt{-1}$, $A$ is the total number of boxes, $X_i$ is the $x$-axis position of the box, and $Y_i$ is the $y$-axis position of the box. The mean value of $Z_{i,j}$ is
\begin{equation}
\overline{Z}_{i,j} = \frac{1}{t_{\rm fin} - t_{\rm ini}} \int_{t_{\rm ini}}^{t_{\rm fin}} Z_{i,j}(t) dt,
\end{equation}
the average of the local phase order parameter of the network is
\begin{equation}
Z_{\rm L} = \frac{1}{625} \sum_{i,j}^{25} \overline{Z}_{i,j}.
\end{equation}

\begin{figure}[htb!]
\centering
\includegraphics[scale=0.29]{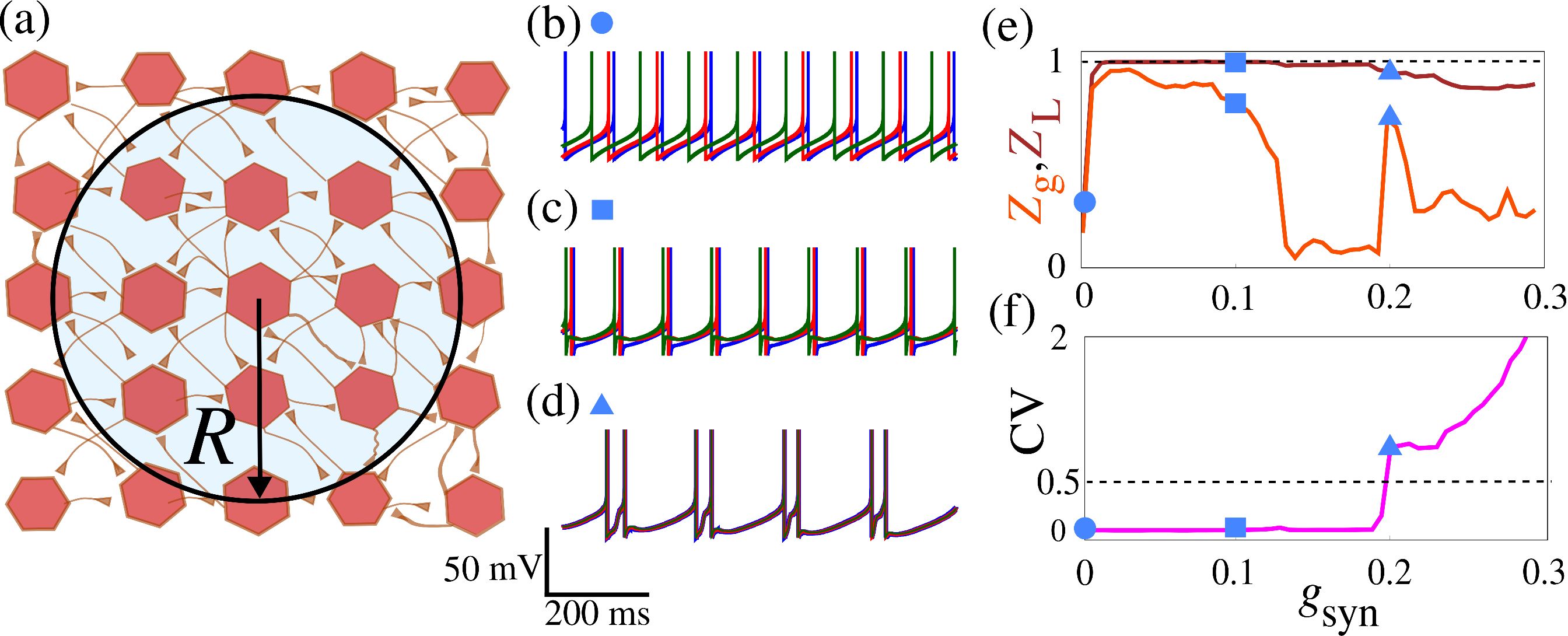}
\caption{The panel (a) displays a schematic representation of the two-dimensional network composed of coupled excitatory neurons. The panels (b-d) exhibit the time evolution of the membrane potential $V$ for three different values of $\it g_{\rm syn}$. We observe (b) desynchronyzed spikes for ${\it g_{\rm syn}} = 0.001$ nS (blue circle), (c) synchronized spike for ${\it g_{\rm syn}} = 0.1$ nS (blue square), and (d) synchronized burst for ${\it g_{\rm syn}} = 0.2$ nS (blue triangle). The panel (e) shows the global and local Kuramoto order parameter as a function of $\it g_{\rm syn}$ and the panel (f) exhibits CV as a function of $g_{\rm syn}$. In panels (b-f), we consider $\it R$ = 70 $\mu$m.}
\label{fig1}
\end{figure}

Figure \ref{fig1}(a) displays an illustrative scheme of the connections in our neuronal network, the neurons connect by means of chemical synapses. There are connections among neurons within the radius {\it R}, otherwise there is no connection. The radius origin is in the neuron position. Figure \ref{fig1} (b-d) exhibit the time evolution of the membrane potential of three different neurons. In panel (b), for a low coupling strength ${\it g_{\rm syn}}=0.001$ nS, the neurons are not synchronized. Increasing the coupling strength to ${\it g_{\rm syn}}=0.1$ nS, the neurons fire in a synchronized spike pattern, as shown in panel (c). The panel (d) shows neurons synchronized and firing according to a burst pattern for ${\it g_{\rm syn} }= 0.2$ nS. The values of the local and global phase order parameters are exhibited in panel (e), where the circles, squares and triangles correspond to the firing patterns of the panels (b-d). The neurons in the network synchronize (locally and globally) for small values of $\it g_{\rm syn}$ and the global order parameter varies much more than the local order parameter. The panel (f) displays the CV as a function of $\it g_{\rm syn}$. The CV value below the dashed line indicates the spike fire patterns. There is a maximum value of coupling strength before the transition from spike to burst behaviors, $g_{\rm syn} = 0.2$ nS.

%%%%%%%%%%%%%%%%%%%%%%%%%%%%%%%%%%%%%%%%%%
%%%%%%%%%%%%%%%%%%%%%%%%%%%%%%%%%%%%%%%%%%
\section{Results}

A bidirectional neuronal network is able to exhibit spiral wave activities \cite{Santos2021} with different amounts of phase singularities (PS) and spiral waves. PS is the spiral core and is composed of asynchronous neurons. Each spiral wave rotates around its own PS and can share the same core. We observe that the local order parameter is below 0.7 in the PS. We define a $\overline{Z}_{i,j}$ threshold to characterize the phase singularities. In this work, we focus on studying the correlation of spiral wave emergence with the network parameter $R$ and $g_{\rm syn}$. We propose a reliable method to distinguish spiral wave from other travelling waves in the cortex. We use the diagnostic tools shown in the previous section.

\begin{figure}[htb!]
\centering
\includegraphics[scale = 0.3]{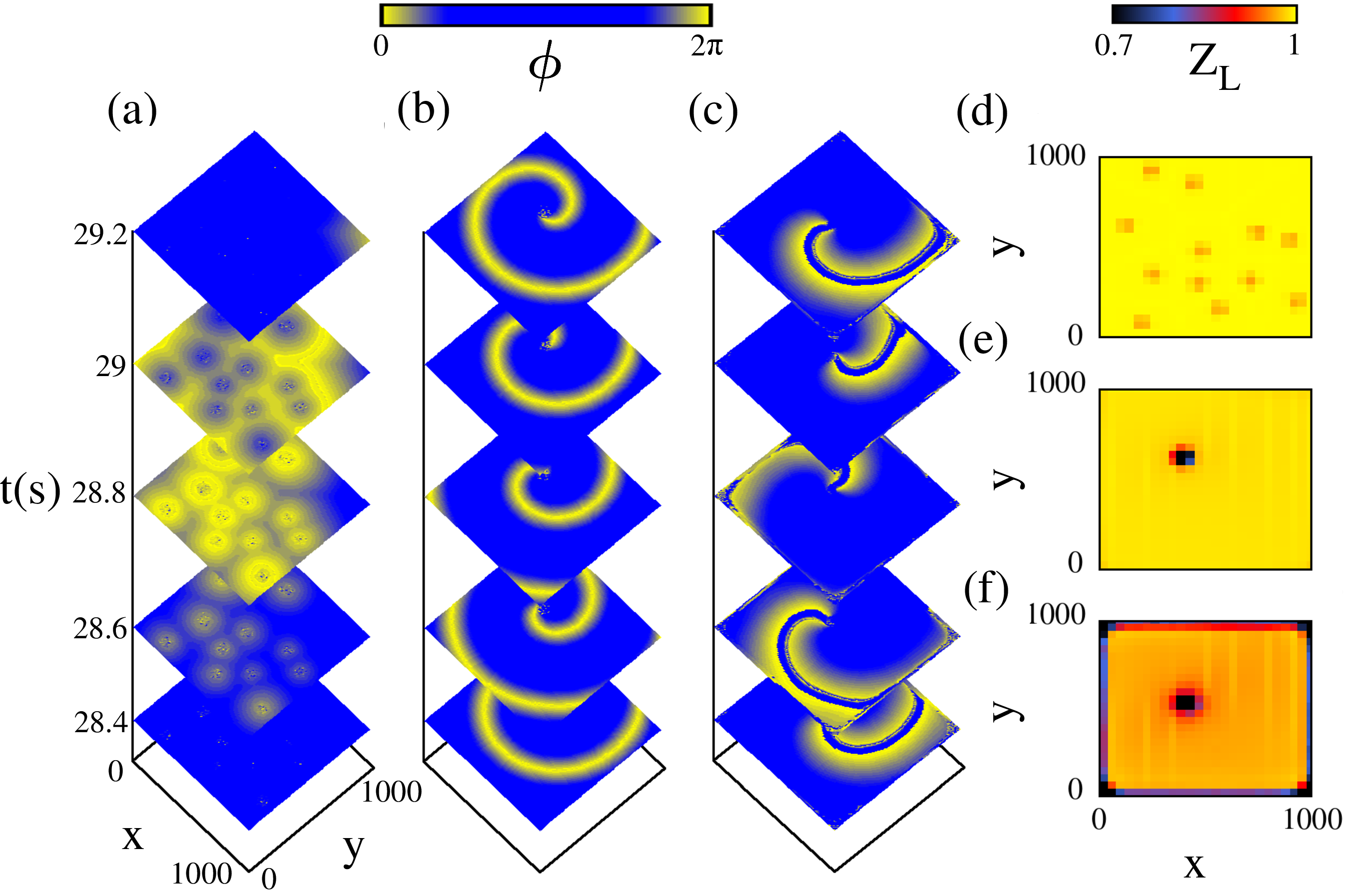}
\caption{(a) Spatial pattern of a synchronyzed spikes (${\it R} = 55$ $\mu$m and ${\it Z_{g}}=0.94$), (b) spiral wave pattern (${\it R}=64.5$ $\mu$m and ${\it Z_{\rm g}}=0.15$), (c) burst spiral wave pattern (${\it R}=75$ $\mu$m and ${\it Z_{\rm g}}=0.36$), (d) local order parameter for  (e) local order parameter of spike spiral wave and (f) local order parameter of burst spiral wave of each neuron $i$. We consider ${\it g_{\rm syn}} = 0.14$ nS.}
\label{fig2}
\end{figure}

Figure \ref{fig2} displays three patterns of waves observable in the network and the local phase order parameter. The panels (a), (b) and (c) exhibit the phase (color bar) of each neuron in a time instant and the panels (d), (e) and (f) show the local order parameter of each wave pattern. The ring waves are shown in Figure \ref{fig2}(a). The wave pattern produces some dark orange dots in the panel (d), which are the origin of where the ring waves propagate from. The dark orange spots are not PS due to the fact that the local order parameter is not below 0.7. The spiral wave is displayed in the panel (b) and its centre is shown in the panel (e), by the black region. In the panel (e), we calculate the local order parameter of the spiral wave, where the black area corresponds to the phase singularity (PS). The mean local order parameter ($Z_{\rm L}$) does not decrease significantly with the presence of the phase singularity, due to the high quantity of local synchronous neurons. For higher $R$ values, it is possible to observe a burst spiral wave, shown in Figure \ref{fig2}(c). There are some border abnormalities in this case. In Figure \ref{fig2}(f), the border has low values of $Z_{\rm L}$, however, we do not count it as PS due the fact that there is no spiral wave rotating around it. To avoid mistakes in the PS counting, we discard the borders. High values of $R$ increase the border effects due to the high connectivity of each neuron, although $R$ influences the firing pattern.

\begin{figure}[htb!]
\centering
\includegraphics[scale=0.28]{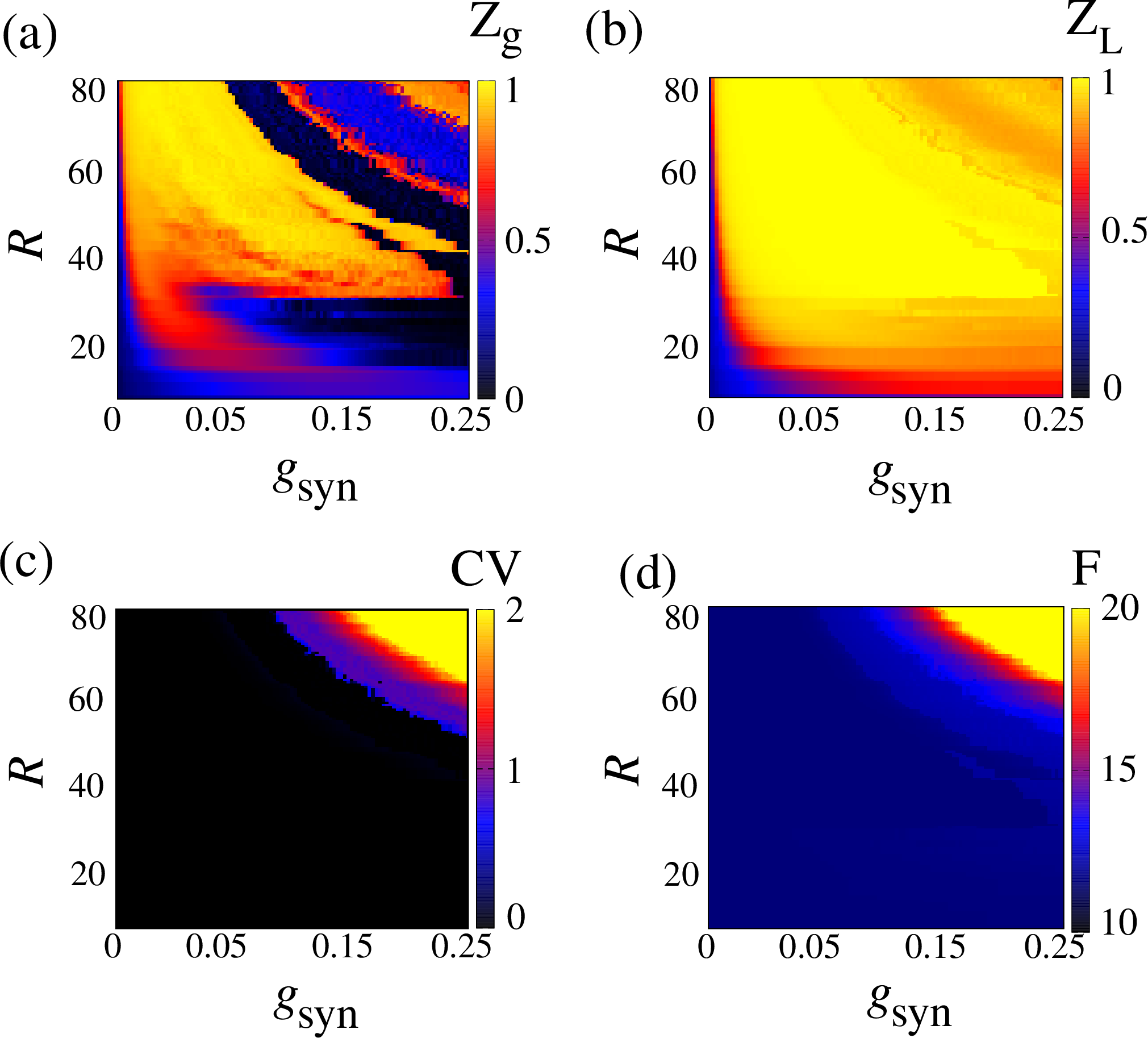}
\caption{(a-d) Parameter space of the network. The color code represents the values of (a) the global order parameter, (b) the local order parameter, (c) the mean coefficient of variation (CV) of the neuronal inter-spike interval and (d) the mean firing rate.}
\label{fig3}
\end{figure}

Figure \ref{fig3} exhibits the parameter space of the four diagnostic tools for $R=75$ $\mu$m. The panel (a) shows the parameter space of the global phase order parameter. The dark regions correspond to the asynchronous activity ($Z_{\rm g} \in [0,0.7]$) and the brighter regions are the synchronized states of the whole network. The network shows synchronous behavior for high values of $R$. The value of $\it g_{\rm syn}$ also plays an important role in the synchronization. Increasing it, we verify transitions from asynchronous to synchronous and vice-versa. The local phase order parameter is displayed in the panel (b). The coupling strength is of great importance for the emergence of synchronous states. The CV suffers changes with the increase of $R$ and $g_{\rm syn}$. It is identified transitions from spike-to-burst, shown in the panel (c). The panel (d) exhibits the average firing rate, there is a quick increase for high values of $R$ and $g_{\rm syn}$. In some regions of high $g_{\rm syn}$, it occurs the synchronization of bursting, which is strongly related to pathological cases, such as epilepsy seizures \cite{Protachevicz2019}. The panels (a) and (b) are quite different in some regions, showing regions in which the network is globally desynchronized meanwhile local neurons are synchronized, $Z_{\rm g}<1$ and $Z_{\rm L} \approx 1$. This result characterizes waves in the network, however, it is not enough to distinguish the wave types.

It is necessary more conditions to characterize spiral waves. The spiral waves have a particular characteristic, that is the phase singularities (the centre which the waves rotate around). However, high amounts of PS are related to asynchronous activity. We establish a maximum value of PS to avoid asynchronous activity. We propose the following conditions to detect spiral waves: (i) the mean global order parameter has to point out asynchronous activity, $Z_{\rm g} \leq 0.7$, (ii) the mean local order parameter has to indicate a local-synchronized activity, $Z_{\rm L} \approx 1$, (iii) there must be at least one phase singularity and PS is characterized by a low local phase order parameter, $\overline{Z}_{i,j}\leq 0.7$, (iv) the number of PS is limited to a maximum number ($P_{\rm max}$), $0 < {\rm PS} \leq P_{\rm max}$. We consider $P_{\rm max} = 20$. 

\begin{figure}[htb!]
\centering
\includegraphics[scale=0.3]{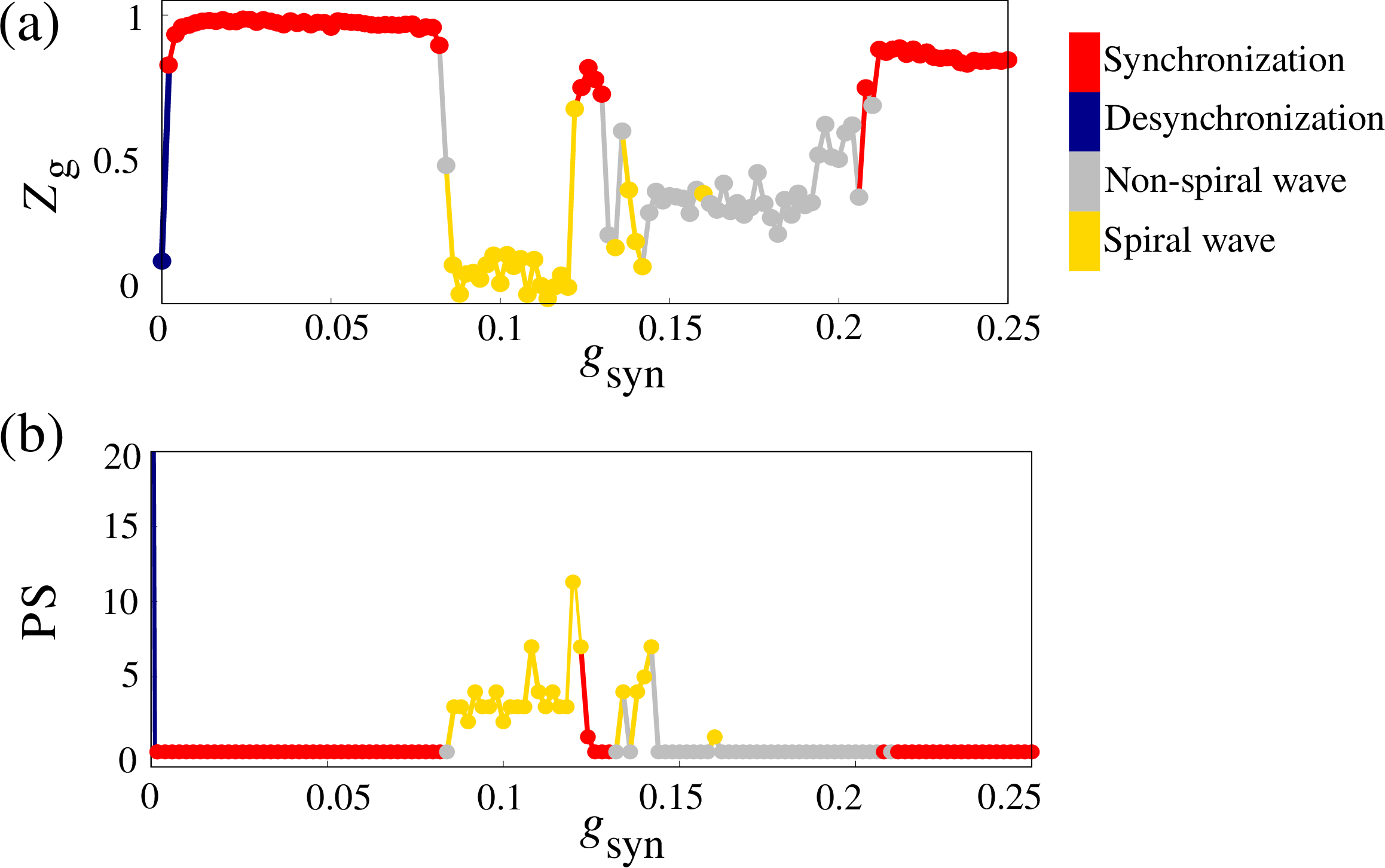}
\caption{In (a) and (b) we plot, respectively, the global phase order parameter and the number of phase singularity as function of $g_{\rm syn}$ for $R = 75 \mu m$. The colors in (a) and (b) represent the dynamical pattern of the network.}
\label{fig4}
\end{figure}

Conditions (i) and (ii) detect wave activity in the network, however, they do not describe the type of wave. The main difference among the wave types is their local phase order parameter pattern. For example, in the ring wave pattern, the local order parameter does not exhibit PS, as shown in Figure \ref{fig2} (d). The presence of PS is characteristic of spiral waves, then at least one focus is necessary, condition (iii). Multi-phase singularities can be observed and the network becomes asynchronous when there are too many ones. We consider a maximum number of PS, condition (iv). These conditions all together indicate spiral wave in the network. The color scheme in Figs. \ref{fig4}(a) and \ref{fig4}(b) show the transition of patterns for the region of $R = 75$ $\mu$m. The yellow points denote the spiral waves, the gray points correspond to the non-spiral waves, the blue points indicate the asynchronous activity and the red points show the synchronized states ($(Z_{\rm g}, Z_{\rm L}) \approx 1$).

\begin{figure}[htb!]
\centering
\includegraphics[scale=0.265]{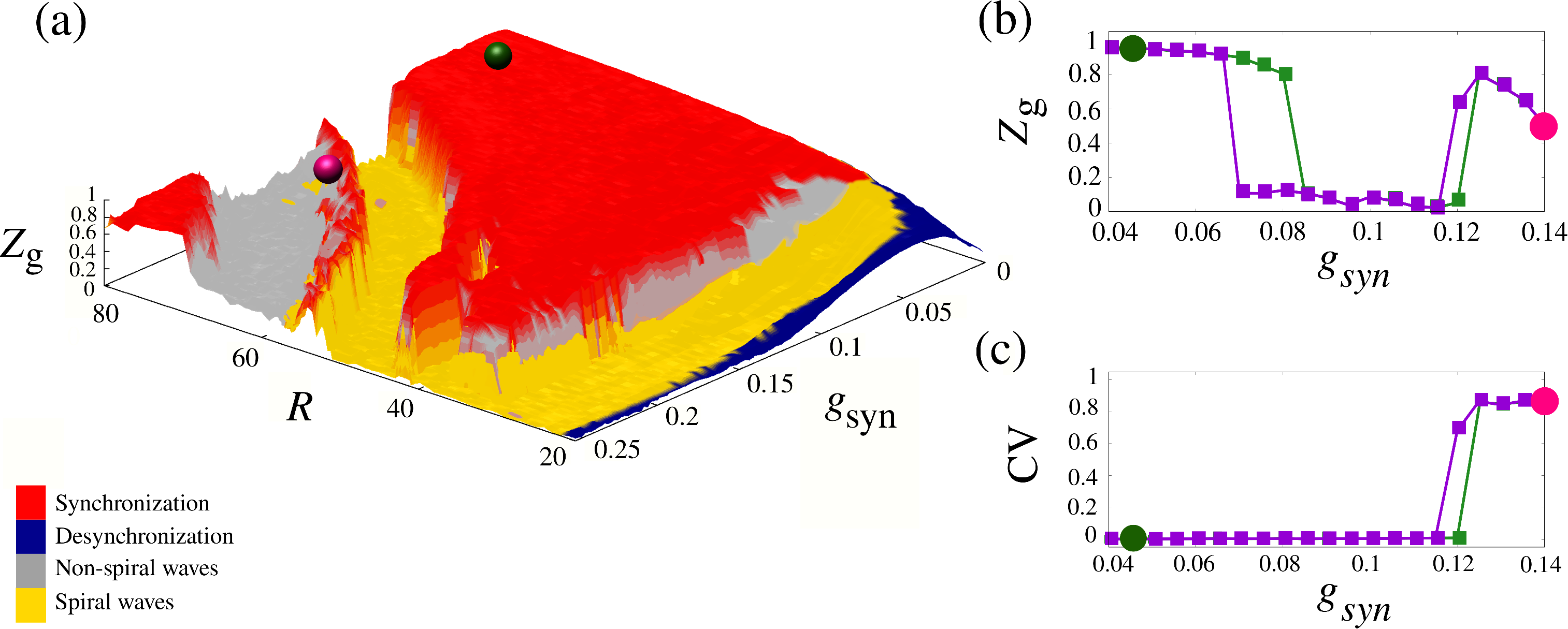}
\caption{(a) Parameter space of the network activity pattern, where each color represents one type of pattern. The yellow region corresponds to the spiral wave, the gray area is related to the non-spiral wave (for example ring pattern), the blue region denotes the desynchronized pattern and the red area corresponds to the synchronous pattern. The panels (b) and (c) show the hysteresis diagram for $R=75$ $\mu$m, (b) global order parameter and (c) coefficient of variation. Asynchronous initial conditions are considered in the purple squares, meanwhile, synchronous initial conditions are used in the green squares. The greenish and pinkish hexagons exhibit the points which start and end the hysteresis cycle, respectively. }
\label{fig5}
\end{figure}

The pattern parameter space of $R\times g_{\rm syn}$ is shown in Figure \ref{fig5}(a), where the colors yellow, gray, blue and red correspond to the spiral waves, non-spiral waves, asynchronous and synchronous patterns, respectively. For $R \leq 20$ $\mu$m, it only generates asynchronous activity. Hence, the radius of connection has great influence in the type of pattern. There are some non-spiral waves regions in the range $40 \leq  R \leq 35$ $\mu$m and $0.25 \leq g_{\rm syn} \leq 0.05$ nS, indicating a coexistence of synchronous and non-spiral waves pattern. It is important to highlight that the radius of connection is about 70 $\mu$m or greater in the human neocortex \cite{Bezaire2016}. The emergence of burst spiral is intrinsically correlated with long range connections ($R \geq 60$ $\mu$m) and intense coupling strength ($g_{\rm syn} \geq 0.12$ nS). The greenish and pinkish hexagons are the starting and ending point of the hysteresis analysis, respectively. Hysteresis diagrams of $Z_{\rm g}$ and CV are displayed in Figs. \ref{fig5}(b) and \ref{fig5}(c), respectively. Asynchronous and synchronous initial conditions are considered in the purple and green squares. The CV bistable region coincides with one of the regions in the $Z_{\rm g}$, both are observed in the $g_{\rm syn}$ range [0.11, 0.13] nS. Two different outcomes are possible, that are burst synchronization for asynchronous initial conditions and spikes desynchronization for synchronous initial conditions. There is one more bistable region in the panel (b), in which the spike desynchronization is easier achieved for synchronous initial conditions. 

%%%%%%%%%%%%%%%%%%%%%%%%%%%%%%%%%%%%%%%%%%
%%%%%%%%%%%%%%%%%%%%%%%%%%%%%%%%%%%%%%%%%%

\section{Conclusions}

In this work, we investigate the emergence of spiral wave patterns within a bidirectional network composed of pyramidal neuron cells. The individual dynamics of each neuron is described by the adaptive exponential integrate-and-fire model. The roles of spiral waves in cognitive activities is not fully understood. In this way, comprehending how network parameters influence the emergence of spirals and developing the capability to detect them is crucial for advancing our understanding of their role in cognitive activities. Hence, we investigate how network parameters influence the emergence of spirals. Additionally, we introduce a novel method capable of identifying chimera spiral waves.

We show that the network composed of coupled pyramidal neurons generates different spatial-temporal patterns. The local and global phase order parameters together have shown to be great tools of wave detection. We observe that wave activity generate a specific range of values for both mean order parameters. Locally the neurons are synchronized for almost all values of $R$ and $g_{\rm syn}$, indicating that waves do not disrupt the local dynamics. The waves greatly affect the global dynamics and makes the global order parameter goes to values close to 0, characterizing desynchronization in the network. The low impact that waves have in local synchronization is due to the firing of the closest neighbors when the travelling wave arrives at the neuron. In the entire network, the passage of wave helps the local neurons to fire meanwhile the neurons that are not at the wave location do not receive a firing support. These characteristics hold great power to identify waves in the network, thus we suggest a diagnostic capable of identifying waves. They are identified by a low global and high local order parameters.  Applying this criteria, we identify some waves. Mostly spike waves are found for radius lower than 55 $\mu$m. The type of wave is not pointed out by this diagnostic, then more criteria to identify spiral waves are necessary.

The identification of spiral waves can not be done by using only the condition proposed earlier, due to the fact that it does not specify the wave type. The spiral waves generate a specific abnormality in the phase
producing very low local order parameter in it. The neurons localized in this abnormality are desynchronized and are the core of the spiral, called phase singularity (PS). The neurons in the PS exhibit a low value in the local order parameter inferior to 0.7, then we propose that PS are characterized by $\rm \overline{Z}_{i,j}<0.7$. It is not necessary only one PS to exist spiral waves, multi-phase singularies are possible. However, a great amount of PSs promote desynchronized activity. We define a maximum number of PSs in the network, $P_{\rm max} = 20$. We propose four conditions which must be met to identify spiral waves, these conditions regard the range of both phase order parameters and the amount of PS. The synchronized states are defined by high global and local order parameters, non-spiral waves are characterized by $(Z_{\rm L} \approx 1$, $Z_{\rm g}<0.7$ and PS = 0, spiral waves exhibit the same conditions of non-spiral waves with the addition of PS $>0$ and desynchronized states are characterized by $Z_{\rm L}\approx 0$ and $Z_{\rm g} \approx 0$. 

The spiral waves are generated by different combinations of $R$ and $g_{\rm syn}$. This wave pattern is not only restricted for spike firing, but also for burst spiral. With regard to the burst spiral pattern, it is not as easily observed as spike spiral. The burst spiral appearance is restricted for specific values of $R$ and coupling strength, for example $R=75$ $\mu$m and $g_{\rm syn}=0.14$ nS. The emergence of non-spiral wave is related to the initial conditions. To understand the initial conditions influence in the system, we analyze the hysteresis diagram for a constant radius and vary the $g_{\rm syn}$. Desynchronous behavior is easier to achieve if the initial conditions are desynchronized. We find a region of bistability in the range $g_{\rm syn}=$[0.11, 0.13] nS.  Depending on the initial conditions, it is possible to verify the existence of spike or burst waves. The identification of bistable state is crucial due to its strong correlation with epileptic seizures \cite{borges2023}. 
%%%%%%%%%%%%%%%%%%%%%%%%%%%%%%%%%%%%%%%%%%
%%%%%%%%%%%%%%%%%%%%%%%%%%%%%%%%%%%%%%%%%%

\section*{Acknowledgments} 

We acknowledge the support from CAPES (Finance Code 88887.849164/ 2023-00; 88881.846051/2023-01; 88881.143103/2017-01), CNPq (403120/2021-7; 301019/2019-3), and FAPESP (Grant Nos. 2018/03211-6; 2020/04624-2; 2022/04251-7).

%%%%%%%%%%%%%%%%%%%%%%%%%%%%%%%%%%%%%%%%%%
%%%%%%%%%%%%%%%%%%%%%%%%%%%%%%%%%%%%%%%%%%

\end{document}